\documentstyle[11pt,newpasp,twoside,epsf]{article}
\def\edcomment#1{\iffalse\marginpar{\raggedright\sl#1\/}\else\relax\fi}
\reversemarginpar

\begin{document}
\title{ Synthesis of small and large scale dynamos }
\author{ Kandaswamy Subramanian }
\affil{ National Centre for Radio Astrophysics, TIFR,
Poona University Campus, Ganeshkhind, Pune 411 007, India. }

\begin{abstract}
Using a closure model for the evolution of magnetic
correlations, we uncover an interesting plausible saturated state of the 
small-scale fluctuation dynamo (SSD) and a novel anology between quantum 
mechanical tunneling and the generation of large-scale fields.  
Large scale fields develop via the $\alpha$-effect, but as
magnetic helicity can only change on a resistive timescale, the time it 
takes to organize the field into large scales increases with magnetic 
Reynolds number. This is very similar to the results which obtain 
from simulations using full MHD.
\end{abstract}

\section{ Fluctuating field dynamics}

The dynamics of the fluctuating magnetic field ${\bf B}$, 
is governed by the induction equation. The velocity is assumed 
to be the sum of a Gaussian random,
delta-correlated in time ${\bf v}_T$, and an ambipolar diffusion type 
component ${\bf v}_D = a[({\bf \nabla } \times {\bf B}) \times {\bf B})]$.
(Here $a = \tau / (4\pi \rho)$, $\tau$ is some response time, and 
$\rho$ is the fluid density).
Assuming that the magnetic field is also Gaussian random, Subramanian 
(1997, 1999; S99) derived closure equations for the longitudinal 
correlation function $M(r,t)$ and the correlation function for 
magnetic helicity density, $N(r,t)$. The random ${\bf v}_T$  
has a longitudinal correlation function $T(r)$ and a correlation function for 
the kinetic helicity density, $C(r)$. 
Defining the operators $\tilde{D}(f)=(1/r^4)\,(\partial (r^4 f)/\partial r)$,
and $D(f)=(\partial f/\partial r)$, we then have, 
\begin{equation}
\dot{M}=2\tilde{D}(\eta_{\rm T}DM)+2GM+4\alpha H; \quad
\dot{N}=-2\eta_{\rm T} H+\alpha M,
\label{closure}
\end{equation}
where $H = - \tilde{D}DN$ is the correlation function of the current
helicity, $G=-\tilde{D}DT$ is the effective induction. Also 
$\alpha=\alpha_0(r) + 4aH(0,t)$ and 
$\eta_{\rm T}=\eta+ \eta_0(r) + 2 a M(0,t)$
are functions resembling the usual $\alpha$-effect and
the total magnetic diffusivity. Here $\alpha_0(r)=-2[C(0)-C(r)]$ and
$\eta_0(r)=T(0)-T(r)$ and $\eta$ the microscopic diffusion. 
Note that at large scales
$r\to \infty$, $ \alpha \to \alpha_{\infty} =
-2C(0) + 4aH(0,t)$ and $\eta_T \to \eta_{\infty} = \eta + T(0) + 2aM(0,t)$.
The $\alpha$-effect suppression is similar to the $\alpha$-suppression 
formula first found by Pouquet et al. (1976); $\eta_T$ 
is however enhanced by the growing field
energy, as in ambipolar diffusion. 

\subsection{ Small-scale dynamo saturation}

The SSD problem, with  $C(r) = 0$, $a=0$
has solutions with $ H(r,t) = 0$, and was first solved by 
Kazantsev (1968). A transformation 
of the form $\Psi(r)\exp(2\Gamma t)  =r^2\sqrt{\eta_T}M(r,t)$,
maps the problem of getting $\Gamma > 0$ modes,
into a bound state problem in time-independent
quantum mechanics (QM). For turbulent motions on a scale $L$,
with a velocity $v$, bound states obtain
provided the magnetic reynolds number (MRN)
$R_m = vL/\eta > R_c \approx 60$, and imply $M$ growing at rate $\sim v/L$.
Further, the bound-state eigen-function describes a 
field, which is strongly concentrated within the diffusive scale, 
$r =r_d \approx L (R_m)^{-1/2}$, and curved on the scale $L$.
For $a\ne 0$, $\eta$ is simply replaced by an effective, time dependent 
diffusion $\eta_D = \eta + 2aM(0,t)$. So as the field (and $M$) grows,
the effective MRN $R_D(t) = v L/ \eta_D(t)$ is driven to the critical
value $R_c$. The final saturated state is obtained 
when $R_D(t_s) = v L/( \eta + 2aM_L(0,t_s) ) = R_c \sim 60$.
So at saturation, the average energy density 
$E_{B}(t_s) = (3M_L(0,t_s) / 8\pi) = (3 / 2)
(\rho v^2 /2) (L/v\tau) (1/ R_c)$.
For $\tau \sim L/v$, $E_B$ is a small fraction $\sim R_C^{-1} \ll 1$,
of the equipartition value. If we intepret the saturated field configuration
in terms of flux ropes with peak field $B_p$, thickness $r_d$,
and curved on scale $L$, $E_B \sim (B_p^2/8\pi) L r_d^2/L^3 $. 
Using $r_d^2/L^2 \approx R_c^{-1}$, and $\tau \sim L/v$, we then have 
$B_p^2/8\pi \sim \rho v^2/2 $, where, remarkably, the $R_C^{-1}$ 
dependence has disappeared.
So the SSD could saturate with the small-scale field 
of equipartition strength, being concentrated into flux ropes of
thickness $LR_c^{-1/2}$, and curved on scale $L$.

\subsection{Large scale dynamo as a tunneling problem}

For helical turbulence with $C(0) \ne 0$, 
new generation terms arise at $r \gg L$, due to
the $\alpha$- effect, in the form
$\dot M = .... +4\alpha_{\infty} H$ and $\dot N = ... + \alpha_{\infty} M$.
These lead to the growth of large-scale correlations on a scale $D$,
with a growth rate $\sim \alpha_{\infty}/D - \eta_{\infty}/D^2$,
as in the large-scale $\alpha^2$- dynamo. A special wavenumber, 
$k_{\rm p}(t) = \alpha_\infty(t)/\eta_\infty(t)$,
is also picked out for any quasi-stationary state. 
For such states, which obtain when one neglects slow resistive
evolution (see below), $(\partial N /\partial t) \approx 0$, implying
$H \approx (\alpha/2\eta_T) M$. And 
if we define $\Psi =r^2\sqrt{\eta_T}M$, 
the equation  $(\partial M/\partial t) \approx 0$, 
can again be mapped to a QM potential problem, for 
the zero-energy eigen-state. However the modified potential now tends 
to a negative definite constant value of 
$ -\alpha_{\infty}^2/\eta_{\infty}$ at large $r$
and so allows tunneling (of the bound state) (see S99).
In fact for $r \gg L$, one has an analytic 
solution $ M(r) \propto  r^{-3/2} J_{3/2}(k_{\rm p} r)$,
exactly as one would get if the large scale field, ${\bf B}_0$, was random
and force-free with ${\bf \nabla} \times {\bf B}_0 = k_{\rm p} {\bf B}_0$.

\section{ Helicity constraint and resistively limited growth}

The closure equations have also been solved numerically
by Brandenburg and Subramanian (2000) (BS2K), adopting forms for
$T(r)$ and $C(r)$ to match closely with the direct
simulation of Brandenburg (2000) (B2000) (Run 5). 
One sees an initial exponential
growth of the magnetic field, which terminates when its
energy becomes comparable to the kinetic energy. 
Note that our closure equations satisfy the helicity
constraint $\dot N(0) = -2 \eta H(0)$.
The numerical solutions show that, after some time $t_{\rm s}$,
the current helicity $\langle {\bf J} \cdot {\bf B} \rangle \propto H(0,t)$,
is driven to a constant value
which however is such that $|\alpha_\infty|$ remains finite.
A constant $H(0,t)$ implies that the magnetic helicity
$\langle {\bf A}\cdot{\bf B}\rangle \propto N(0,t)$ grows linearly
at a rate proportional to $\eta$.
During this phase the magnetic field correlations
can extend to larger and larger
scales. The corresponding magnetic energy spectra,
$E_{\rm M}(k,t)=(1/\pi)\int_0^{\infty} M(r,t)\,(kr)^3\,j_1(kr)\,dk$\, 
are shown in Figure 1.

\epsfxsize=11.2cm
\begin{figure}[t]\epsfbox{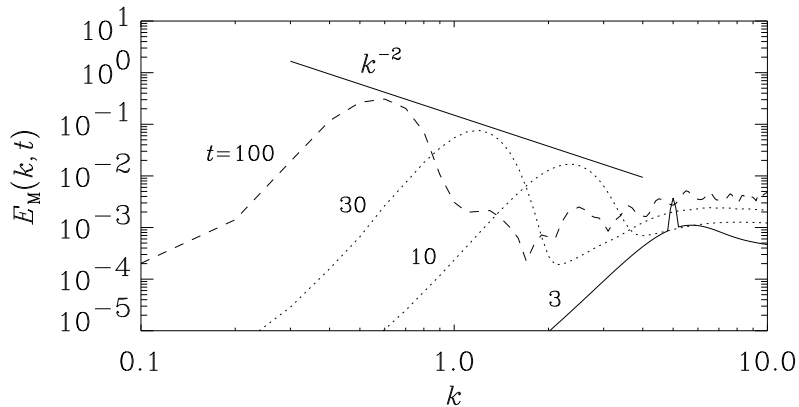}\caption[]{
Evolution of magnetic energy spectra.
Note the propagation of magnetic helicity and energy to progressively
larger scales. The $k^{-2}$ slope is given for orientation.
(see BS2K)
}
\end{figure}

The resulting magnetic field is strongly helical (cf. section 1.2)
and the magnetic helicity spectra (not shown) satisfy 
$|H_{\rm M}|\la(2/k)E_{\rm M}$. The
development of a helicity wave travelling towards smaller and smaller
$k$, as seen in Figure 1, is in agreement with the closure model
of Pouquet et al. (1976). We have also checked that to a
very good approximation the wavenumber of the peak is given by
$k_{\rm peak}(t)\approx k_{p}(t)$, as expected from section 1.2, 
and it decreases with time because $\alpha_\infty$ tends to a
finite limit and $\eta_\infty$ increases.
Further, since the large scale field is helical, and since
most of the magnetic energy is by now (after $t=t_{\rm s}$) in the large
scales, the magnetic energy is proportional to
$\langle {\bf B}^2 \rangle \approx 
k_{\rm p} \langle {\bf A}\cdot{\bf B} \rangle $, and can
therefore only continue to grow at a resistively limited rate.
These results are analogous to the full MHD case (B2000);
the helicity constraint is independent of the nature of the feedback!
In conclusion, our closure model with ambipolar diffusion type
non-linearity provides a useful model, enabling 
progress to be made in understanding nonlinear dynamos. 
One now needs to think of ways, to relax the helicity constraint,
(cf. Blackman and Field (2000), Kleeorin et al. (2000)),
so that large-scale magnetic fields can grow fast enough.

\end{document}